\documentclass[pra,twocolumn,superscriptaddress,amsmath,amssymb,floatfix]{revtex4}

\usepackage{amsmath,amssymb}
\usepackage{graphicx,color}
\usepackage{booktabs}
\usepackage{verbatim}

\usepackage[utf8]{inputenc}

\usepackage{times}
\usepackage{mathptmx}
\DeclareSymbolFont{letters}{OML}{txmi}{m}{it}

\newcommand{\ket}[1]{|#1\rangle}
\newcommand{\bra}[1]{\langle#1|}
\newcommand{\braket}[1]{\langle#1\rangle}
\newcommand{\rmd}{\mathrm{d}}
\newcommand{\rmi}{\mathrm{i}}

\renewcommand{\Re}{\operatorname{Re}}
\newcommand{\TC}{\operatorname{TC}}
\newcommand{\ddt}{\frac{\rmd}{\rmd t}}

\newcommand{\boldpsi}{\psi\!\!\!\!\!\psi}

\newcommand{\tr}{\operatorname{tr}}

\newcommand{\brrakket}[1]{\langle\!\!\langle#1\rangle\!\!\rangle}

\begin{document}

\title{Multiconfigurational time-dependent Hartree method for describing
  particle loss due to absorbing boundary conditions}

\author{Simen Kvaal} 
\email{simen.kvaal@cma.uio.no}
\affiliation{Matematisches Institut, Universität Tübingen, Auf der
  Morgenstelle 10, D-72076 Tübingen, Germany}
\affiliation{Centre of Mathematics for Applications, University of
  Oslo, N-0316 Oslo, Norway} 


\begin{abstract}
  Absorbing boundary conditions in the form of a complex absorbing
  potential are routinely introduced in the Schrödinger equation to
  limit the computational domain or to study reactive scattering
  events using the multi-configurational time-dependent Hartree method
  (MCTDH). However, it is known that a pure wave-function description
  does not allow the modeling and propagation of the remnants of a
  system of which some parts are removed by the absorbing boundary. It
  was recently shown [S.~Selstø and S.~Kvaal, J.~Phys.~B:
  At.~Mol.~Opt.~Phys.~{\bfseries 43} (2010), 065004] that a master
  equation of Lindblad form was necessary for such a description. We
  formulate a multiconfigurational time-dependent Hartree method for
  this master equation, usable for any quantum system composed of any
  mixture of species. The formulation is a strict generalization of
  pure-state propagation using standard MCTDH. We demonstrate the
  formulation with a numerical experiment.
\end{abstract}

\maketitle

\section{Introduction}
\label{sec:introduction}

Today, the \emph{de facto} standard approach in \emph{ab initio}
quantum-mechanical many-particle propagation is the multiconfigurational
time-dependent Hartree method (MCTDH) and its variations
\cite{Meyer1990,Beck2000,Raab2000,Zanghellini2003,Alon2007,Alon2008,Vendrell2011}. Already
for $N=2$ electrons in three dimensions, the full six-dimensional
time-dependent Schrödinger equation is very hard to solve and can only be handled
on supercomputers. With MCTDH for identical particles the exponential scaling
of the Hilbert space dimension with respect to $N$ is ``postponed'' to higher
number of particles, and the $N=2$ propagation can be done on a single desktop
computer. Current implementations can handle $N\leq 8$ electrons in
cylindrical geometries reliably \cite{Nest2005,Caillat2005}. For bosons, the
Pauli exclusion principle is absent, and MCTDH can treat hundreds
\cite{Sakmann2009} and even thousands of particles \cite{Streltsov2009b} in
one-dimensional geometries, and recent multi-layer MCTDH techniques allow
\emph{distinguishable} dimensions in the \emph{thousands} with relative ease
\cite{Vendrell2011}, showing great promise of extending the domain of
application of MCTDH methods.

MCTDH is derived using the time-dependent variational principle
\cite{Lubich2005,Broeckhove1988}. As such, it is energy conserving,
unitary and quasi-optimal in the sense that the growth of the error in
the 2-norm is locally minimized.

\emph{Ab initio} dynamical problems in quantum mechanics are formulated on an
infinite domain which must be truncated for numerical calculations. The
numerical reflections implied by the truncations are usually dealt with using
absorbing boundary conditions of some sort. The most common approach is to
introduce a complex absorbing potential (CAP) in a region around the truncated
domain \cite{Kosloff1986}. That is, the Hamiltonian $H$ is mapped to $H - \rmi
\Gamma$, where $\Gamma\geq 0$ is a local one-body potential vanishing on the
domain of interest, and only taking nonzero values outside the domain. This
approach is also used in order to calculate properties like reaction and
ionization probabilities, and CAPs are routinely implemented in MCTDH
codes~\cite{Zanghellini2003,Caillat2005,Jackle1996,Muga2004,Meyer2009}.

Other absorbing operators are also common, such as the so-called
transformative CAP (TCAP) \cite{Riss1998}, which is more or less equivalent with the non-local
CAP obtained using smooth exterior scaling \cite{Moiseyev1998} or perfectly matched
layers \cite{Nissen2010}. While \emph{exact} and space-local absorbing boundary
conditions may be formulated \cite{Arnold1998}, they are in general non-local
in time, and therefore impractical. In this work, we focus solely on a local
CAP for simplicity, but any absorbing operator can be used.

Given a system of $N$ particles, the wave-function $\Psi_N$ is normalized to the
probability of finding \emph{all} particles within the computational
domain. With a CAP, $\Psi_N$ evolves according to the non-Hermitian Schrödinger equation
\begin{equation}
  \rmi \ddt \Psi_N = (H - \rmi \Gamma) \Psi_N.
  \label{eq:nh-se}
\end{equation}
An elementary calculation gives $\ddt \|\Psi_N\|^2 = - 2
\braket{\Psi_N|\Gamma|\Psi_N}$ for the probability derivative. Consequently,
if the wave-function overlaps the CAP, the whole wave-function decays and
eventually vanishes; it does \emph{not} approach a wave-function with a
different number of particles. In other words, even with an absorbing
boundary, one is stuck with an $N$-particle description. Information like
ionization probabilities and reaction rates may be obtained from evolving
Eq.~(\ref{eq:nh-se}) alone, but if the \emph{remainder} of the system is
desired, i.e., a description of the $N-1$, $N-2$ particle systems, etc, one is
at loss.

In a recent article it was argued that the solution is a density
operator approach \cite{Selsto2010} because the loss of particles is
an irreversible process; $H-\rmi \Gamma$ is a non-Hermitian operator
implying a preferred direction of time. The necessity of the quantum
dynamical semigroup describing the evolution to be trace-preserving,
Markovian and completely positive implies the applicability of the
famous theorems due to Lindblad and Gorini and coworkers
\cite{Lindblad1976,Gorini1976,Alicki2007}, giving a master equation on
Lindblad form.
The resulting equation is 
\begin{equation}
  \ddt \rho_n = -\rmi[H,\rho_n] - \{ \Gamma, \rho_n \} + 2 \int \Gamma(x)
  \boldpsi(x)\rho_{n+1} \boldpsi(x)^\dag \; \rmd x,
  \label{eq:fundamental-master}
\end{equation}
where $\rho_n$ is the density operator for the $n$-particle subsystem, $0 \leq
n \leq N$. The integral is over all discrete and continous degrees of
freedom. It should be noted that the non-Hermitian Schrödinger
equation~(\ref{eq:nh-se}) is equivalent to the von Neumann equation
\begin{equation*}
  \ddt \rho_n = -\rmi[H,\rho_n] - \{ \Gamma, \rho_n \}
\end{equation*}
for each $n$-particle system. 
The original formulation of the $N$-particle problem is not changed, but
extended to yield $n\leq N$ particle systems as a by-product. Importantly,
nowhere is an \emph{ad hoc} hypothesis introduced.

In this article, we formulate a MCTDH method for
Eq.~(\ref{eq:fundamental-master}) for identical particles and
mixtures. It is based on a so-called type II density operator manifold
\cite{Raab1999,Raab2000} and is fully variational. It turns out to be exactly
trace-preserving and a strict generalization of MCTDH for pure states
evolving according to the non-Hermitian Schrödinger equation
(\ref{eq:nh-se}). The method is first formulated for identical
fermions and bosons, and then extended to arbitrary mixtures of
species. The derivation for identical particles is done in
Sec.~\ref{sec:mctdh-for-fermions}, while the generalization is done in 
Sec.~\ref{sec:bosons-and-mixtures}, after a numerical experiment on
a system of identical fermions is presented in
Sec.~\ref{sec:experiments}. Not surprisingly, the resulting
formulation is a direct generalization of the pure state MCTDH treatment
of mixtures \cite{Alon2007b}.

To prepare for the density operator MCTDH method (hereafter called
$\rho$-MCTDH as opposed to $\Psi$-MCTDH for pure states), we give a brief
derivation of Eq.~(\ref{eq:fundamental-master}) in
Sec.~\ref{sec:master-equation} that underlines the inevitability and
uniqueness of the master equation on Lindblad form. We also show how the
probability interpretation of $\|\Psi_N\|^2$, which is not purely quantum
mechanical, can be interpreted in terms of measurements performed continuously
in time.

\section{Lindblad equation for systems with a CAP}
\label{sec:master-equation}

\subsection{Classical probabilities from a CAP}
\label{sec:classical}

The evolution of $\Psi_N$ under the non-Hermitian Schrödinger
equation~(\ref{eq:nh-se}) is irreversible. It introduces a preferred
direction in time, which is easily seen from the fact that the
evolution cannot be reversed: eventually $\|\Psi_N\|>1$, and the
backward propagation may be non-existent, even mathematically.

We now give an interpretation of the square norm $\|\Psi_N\|^2$ in terms of
measurements performed continuously in time, thereby exposing the
irreversibility. To this end, consider first a closed single-particle system,
described by the Hamiltonian $H$---without a CAP---in the Hilbert space
$\mathcal{H}_1$. Quantum mechanically, the squared norm $\|\Psi_1\|^2$ of the
wave-function is the probability of finding the particle \emph{somewhere} in
configuration space. Equivalently, it is the probability of obtaining the
value $1$ upon measurement of trivial observable $I$ (the identity operator).

Adding a CAP $-\rmi\Gamma$ to the Hamiltonian and keeping the
probability interpretation of $\|\Psi_1\|^2$ has non-trivial
implications. The description is obviously no longer in accordance
with the basic postulates of quantum mechanics, since
\begin{equation*}
 \ddt \|\Psi_1\|^2 = -2\braket{\Psi_1|\Gamma|\Psi_1} \leq 0,
\end{equation*}
that is to say, total probability is \emph{not conserved}.

Suppose we perform a single measurement on the observable $P$ given by
\begin{equation*}
  P = \int_\Omega \ket{x}\bra{x} \rmd x ,
\end{equation*}
i.e., of the projection operator
$P$ onto $\Omega$, the truncated computational domain. It has two eigenvalues, $0$ and $1$,
corresponding to finding the particle outside or inside $\Omega$,
respectively. We will obtain the answer $0$ with probability
$1-\|\Psi_1\|_\Omega^2 = 1 - \braket{\Psi_1|P|\Psi_1}$ and the answer $1$ with
probability $\|\Psi_1\|^2_\Omega$. After measurement, the wave-function
collapses onto $(I-P)\Psi_1$ in the former case, and $P\Psi_1$ in the
latter. Note that the wave-function collapse is irreversible, as the
original wave-function cannot be reconstructed after the event.

Suppose we perform many experiments at short time intervals $t=n\tau$,
$n=0,1,2,\ldots$. Each experiment yields certain information about
the system after the measurement: we know with certainty if the particle is in
$\Omega$ or not. After $n$ experiments, all giving $1$ as answer, the wave
function is (to first order in $\tau$)
\begin{equation*}
  \Psi_1(n\tau) = [P e^{-\rmi\tau H}]^n \Psi_1(0), 
\end{equation*}
and the probability that $n$ positive answers have been given is
\begin{equation*}
  \|\Psi_1(n\tau)\|^2,
\end{equation*}
again to first order. This probability is classical in the sense that
it is a probability description of the history of our macroscopic
measurement device, or of its print-out on a sheet of paper if one
prefers. 

If we allow the approximation
\begin{equation*}
  P \approx \exp(-\tau\Gamma), 
\end{equation*}
which dampens $\Psi_1$ strongly outside $\Omega$, we get
\begin{equation*}
  P e^{-\rmi \tau H} \approx e^{-\tau \Gamma} e^{-\rmi\tau H} = e^{-\rmi \tau
  (H - \rmi\Gamma)} + \mathcal{O}(\tau^2).
\end{equation*}
In the limit of small $\tau$, we see that the solution to the Schrödinger
equation with a CAP is obtained. $\|\Psi_1(t)\|^2$ is then the probability of
finding the measurement apparatus in the state ``the particle has not yet been
found outside $\Omega$'' at time $t$.

Our discussion is immediately suggestive of interpreting the CAP as an
actual model for some external detecting device, which was already
pointed out in the first paper on CAPs in quantum mechanics
\cite{Kosloff1986}. Although it is likely that
Eq.~(\ref{eq:fundamental-master}) can be derived in such a way,
it is not relevant here, as the Schrödinger equation with a CAP is our
starting point. The above discussion is only intended as a
means for understanding the irreversibility of the non-Hermitian
dynamics. Whatever interpretation we use, we see that the
single-particle system undergoes a transition to a zero-particle state
(i.e., zero particles in $\Omega$) in an irreversible way, and that the
probability of having $n=1$ or $n=0$ particles is not quantum mechanical.

A complete specification of the quantum state
of the $n\leq 1$-particle system is a density operator in the Hilbert
space $\mathcal{H}_1\oplus \mathcal{H}_0$, where $\mathcal{H}_n$ is
the Hilbert space of $n$ particles. In this case, the density operator
is, on block form,
\begin{equation*}
  \rho = \begin{pmatrix} \ket{\Psi_1}\bra{\Psi_1} & 0 \\ 0 & (1 -
    \braket{\Psi_1|\Psi_1})
  \end{pmatrix}.
\end{equation*}
Note that $\mathcal{H}_0$ is one-dimensional, and the lower right
block is just a non-negative real number, the
probability of zero particles in $\Omega$.

For a rigorous treatment of the above limiting process, see Chapter
7.4 in Ref.~\cite{Davies1976}.  

Generalizing the discussion to $N$
particles is completely analogous to our discussion, approximating
$P\approx \exp(-\rmi\Gamma)$ in the same way. $\|\Psi_N\|^2$ is then
the probability of the measurement apparatus showing that all
particles are inside $\Omega$.

\subsection{$N$-fermion systems with a CAP}
\label{sec:N-fermion-cap}

Suppose a complete set of fermionic or bosonic creation operators
$\{c_j^\dag\}$, $j=1,2,\ldots$, are given. An important formal property of
$c_j^\dag$ is that it defines a map from $\mathcal{H}_n$ to
$\mathcal{H}_{n+1}$, the Hilbert spaces of $n$ and $n+1$ fermions,
respectively, and that they fulfill the (anti-)commutator relation
\begin{equation}
  \{ c^\dag_j, c_k \}_\pm \equiv c^\dag_j c_k \pm c_k c^\dag_j = \delta_{jk}
  \label{eq:fermion-anticomm}
\end{equation}
with the plus sign for fermions, and the minus sign for bosons.

For us, it is natural to work within Fock space $\mathcal{H}$,
defined by
\begin{equation*}
  \mathcal{H} \equiv \bigoplus_{n=0}^\infty \mathcal{H}_n,
\end{equation*}
containing all possible states with any number of particles. As
operators in $\mathcal{H}$, the $c_j$ are orthogonal in the Hilbert-Schmidt
inner product,
\begin{equation*}
  \brrakket{c_j, c_k} = \tr(c^\dag_j c_k) = 0, \quad j\neq k,
\end{equation*}
and traceless,
\begin{equation*}
  \tr(c_j) = 0.
\end{equation*}

Given an $n$-particle Hamiltonian on first-quantized form, viz,
\begin{equation*}
  H_{n} = \sum_{i=1}^n h(i) + \sum_{i=1}^{n-1} \sum_{j=i+1}^n u(i,j),
\end{equation*}
(where the indices $i$ and $j$ in the operators indicate which particles'
degrees of freedom they act on,) we may write it compactly on $n$-independent
form using the expression
\begin{equation}
  H = \sum_{jk} h_{jk} c^\dag_j c_k +
  \frac{1}{2}\sum_{jklm} u_{jklm}
  c^\dag_jc^\dag_k c_m c_l,
  \label{eq:ham-second-quant}
\end{equation}
where $h_{jk}$ and $u_{jklm}$ are the usual one- and two-particle integrals,
respectively.

Adding a cap $\Gamma = \sum_{i=1}^n \Gamma(i)$ amounts to modifying the
single-particle coefficients since
\begin{equation}
  \Gamma = \sum_{jk} \Gamma_{jk} c^\dag_j c_k.
  \label{eq:ham-with-cap}
\end{equation}
It is important to note, that given the set $\{c_j^\dag\}$, the
second-quantized expressions of $H$ and $\Gamma$ are unique, and vice versa.

\subsection{The Lindblad equation}
\label{sec:the-master-equation}

We now discuss the master equation (\ref{eq:fundamental-master}),
introduced in Ref.~\cite{Selsto2010}, for a system of identical
particles. We underline that the master equation \emph{follows
  uniquely} from the probability interpretation of $\|\Psi_n\|^2$ for
an $n$-particle wave-function, and the requirement of a Markovian,
trace-preserving and completely positive evolution. We also address
some mathematical points omitted in Ref.~\cite{Selsto2010}.

Describing a system with a variable number of particles, we work
with states in Fock space. Our starting point, the non-Hermitian Schrödinger
equation with a given CAP, then reads
\begin{equation}
  \rmi \ddt \Psi = (H - \rmi \Gamma)\Psi.
  \label{eq:nh-se-fock}
\end{equation}
With an initial condition with exactly $N$ particles, i.e. $\Psi(0) = \Psi_N
\in \mathcal{H}_N$, this equation is equivalent to
Eq.~(\ref{eq:nh-se}), which we repeat here for convenience:
\begin{equation}
  \rmi \ddt \Psi_N = (H - \rmi \Gamma)\Psi_N.
  \label{eq:nh-se-fock-n}
\end{equation}
This description is valid for all particle
numbers $N$.

\begin{figure}
  \includegraphics{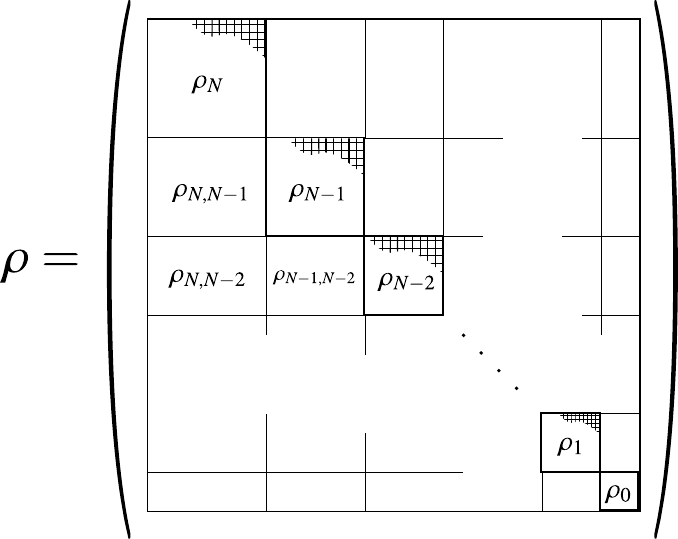}
  \caption{Illustration of the block structure of the density operator in
    Fock space\label{fig:blocks} with at most $N$ identical particles. In general, each
    block is infinite  dimensional (except for $\rho_0$ which is
    $1\times 1$). The diagonal blocks are emphasized, as for pure
    state initial conditions $\rho=\ket{\Psi_N}\bra{\Psi_N}$ the
    density operator turns out to be block diagonal.}
\end{figure}

As Eq.~(\ref{eq:nh-se-fock}) is irreversible, it is not possible to find
some new Hamiltonian $H'$ in Fock space that generates a unitary evolution which 
describes a decreasing number of particles and also reproduces
Eq.~(\ref{eq:nh-se-fock-n}). Instead, one must turn to a 
density operator description. For a density operator $\rho \in
\TC(\mathcal{H})$, the trace-class operators in Fock space \cite{Alicki2007,Davies1976},
Eq.~(\ref{eq:nh-se-fock}) is equivalent to the non-trace conserving von
Neumann-equation
\begin{equation}
  \ddt \rho = -\rmi[H,\rho] - \{ \Gamma, \rho \},
  \label{eq:von-neumann-fock}
\end{equation}
which is verified by a simple computation.
The density operator $\rho$ has a natural block structure. Letting $P_n$
be the orthogonal projector onto $\mathcal{H}_n$, we have the resolution of the identity
\begin{equation*}
  I = \sum_{n=0}^\infty P_n.
\end{equation*}
Applying this to either side of $\rho$ gives the
decomposition
\begin{equation*}
  \rho = \sum_{n=0}^\infty \sum_{m=0}^\infty \rho_{n,m}, \quad \rho_{n,m} \equiv P_n\rho P_m.
\end{equation*}
This block structure is depicted in
Fig.~\ref{fig:blocks}. Projecting
Eq.~(\ref{eq:von-neumann-fock}) from the left and right with
$P_n$, we obtain
\begin{equation}
  \ddt \rho_{n,n} = -\rmi[H,\rho_{n,n}] - \{ \Gamma, \rho_{n,n} \},
  \label{eq:von-neumann-fock-n}
\end{equation}
which is equivalent to Eq.~(\ref{eq:nh-se-fock-n}).

We wish obtain a master equation for $\rho$ that \emph{does} preserve total
probability. At all times, there is a probability $\tr(\rho_n)$ of
having $n$ particles (we define $\rho_n\equiv\rho_{n,n}$), and these should
add up to $\tr(\rho)=\sum_n\tr(\rho_n) = 1$. Since the von-Neumann equation is
local in time, we require the master equation to be Markovian. Our master
equation must reproduce~(\ref{eq:von-neumann-fock-n}) for $n=N$, where $N$ is
the number of particles initially present. Of course, for $n<N$ the master
equation will end up with some source terms. This must hold for all $N$, i.e.,
our master equation should not depend on $N$; only the initial condition
depends on $N$.

We also require the master equation to describe a physical system which in turn
may interact with other quantum systems. A minimal requirement for such master
equations is that its quantum semigroup is \emph{completely positive}. Complete positivity
roughly states that if our system is described together with a different system with
Hilbert space $\mathcal{V}$, so that
the combined state is $\sigma\in \TC(\mathcal{H}\otimes \mathcal{V})$, then the semigroup
of $\sigma$ preserves the self-adjoint positive-semidefiniteness of
$\sigma$. Surprisingly, requiring this property on the flow of $\rho$ alone is not enough
\cite{Gorini1976,Alicki2007}. 

Let us for the moment assume that Fock space has \emph{finite
  dimension}, i.e., it is defined by finite number $L$ of creation
operators. The theorem by Gorini and coworkers is now applicable
\cite{Gorini1976}: Any trace-preserving, Markovian and completely
positive quantum dynamical semigroup has a master equation on the form
\begin{equation}
  \ddt{\rho} = -\rmi [H, \rho] + \mathcal{D}(\rho),
  \label{eq:generic-master-equation-repeat}
\end{equation}
where the dissipative terms have the generic form
\begin{equation}
  \mathcal{D}(\rho) = \sum_{\alpha\beta} d_{\alpha\beta} ( 2 L_\beta \rho
  L_\alpha^\dag - \{ L_\alpha^\dag
  L_\beta, \rho \}).
  \label{eq:dissipative-term-repeat}
\end{equation}
The operators $L_\alpha$ form a traceless, orthogonal set with the
Hilbert-Schmidt inner product $\brrakket{\rho,\sigma} =
\tr(\rho^\dag\sigma)$ (the operators are then linearly independent in
$\TC(\mathcal{H})$), and $[d_{\alpha\beta}]$ is a Hermitian positive
semi-definite matrix. Importantly, the quantum dynamical semigroup
generated by Eq.~(\ref{eq:generic-master-equation-repeat}) is
uniquely given by $H$ and $\mathcal{D}$ and vice versa. 

The $L_\alpha$ play the rôle of a basis in $\TC(\mathcal{H})$, which is of
dimension $\dim(\mathcal{H})^2$. It is of course not unique, but given
a choice of basis, the coefficients $[d_{\alpha\beta}]$ are unique.
Notice that the trace of the anti-commutator term in
Eq.~(\ref{eq:dissipative-term-repeat}) is equal to but of opposite
magnitude compared to the term where $\rho$ is sandwiched between
$L_\beta$ and $L_\alpha^\dag$.

The von-Neumann equation is \emph{almost} of the
form (\ref{eq:generic-master-equation-repeat}); it lacks the ``sandwiched
terms'' responsible for a compensation of the trace decrease due to the
anti-commutator, i.e.,
\begin{equation*}
  \ddt\rho = -\rmi [H,\rho] - \sum_{jk} \Gamma_{jk} \{ c^\dag_j c_k, \rho \}.
\end{equation*}
As the operators $\{c_j\}$ are indeed traceless
and orthogonal, it is immediately clear that the correct master
equation is obtained by simply adding the sandwiched terms, i.e.,
\begin{equation}
  \ddt\rho = -\rmi [H,\rho] - \sum_{jk} \Gamma_{jk} \{ c^\dag_j c_k, \rho \} +
  2\sum_{jk} \Gamma_{jk} c_k \rho c^\dag_j.
  \label{eq:master-eq-second-quant}
\end{equation}
If the dissipative terms are not chosen on \emph{exactly} this form,
Eq.~(\ref{eq:von-neumann-fock-n}) will not be reproduced for all
$N$.

Indeed, consider an initial condition on the form
\begin{equation}
  \rho(0) = \ket{\Psi_N}\bra{\Psi_N}.
  \label{eq:pure-state-ic}
\end{equation}
Projecting Eq.~(\ref{eq:master-eq-second-quant}) from the left and right
with $P_n$ and $P_m$, respectively, we obtain the differential equation obeyed
by the block $\rho_{n,m}$,
\begin{equation}
  \ddt \rho_{n,m} = - \rmi [H, \rho_{n,m} ] - \{ \Gamma, \rho_{n,m} \} + 2
  \sum_{jk} \Gamma_{jk} c_k \rho_{n+1,m+1} c^\dag_j.
  \label{eq:master-eq-second-quant-blocks}
\end{equation}
We see that the off-diagonal blocks $\rho_{n,m}$ with $n\neq m$ are
identically zero with the initial
condition~(\ref{eq:pure-state-ic}). The compensating sandwich terms
are seen to be responsible for transporting probability from the
$N$-particle system into the $N-1$ particle system, and so on,
downwards along the diagonal blocks in Fig.~\ref{fig:blocks}. Moreover, the
evolution of $\rho_{N,N}$ is equivalent to
Eq.~(\ref{eq:nh-se-fock-n}). Modifying the coefficients $\Gamma_{jk}$ or
adding more linearly independent operators to the set $\{c_j\}$ will void this
for some $N$.

The theorem by Gorini and coworkers is only valid in a finite
dimensional Hilbert space. For infinite dimensional spaces, Lindblad
discovered a generalization for norm-continuous semigroups
\cite{Lindblad1976}, for which $H$ and $\Gamma$ are necessarily
bounded operators. This is rarely the case, but a Lindblad-type theorem
for the general unbounded cases is simply not known
\cite{Alicki2007}. On the other hand, all known examples of completely
positive semigroups in infinite dimensional spaces have generators on the generic form
(\ref{eq:dissipative-term-repeat}), with possibly unbounded $H$ or
$[d_{\alpha\beta}]$.

It now seems reasonable to remove the restriction of a finite-dimensional Fock
space, which does not change the formal appearance of the Lindblad equation
(\ref{eq:master-eq-second-quant}). Using the relation
\begin{equation*}
  \boldpsi(x)^\dag \equiv \sum_j \overline{\varphi_j(x)} c_j^\dag,
\end{equation*}
where $\{\varphi_j(x)\}$ is the orthonormal single-particle basis functions associated
with $\{c_j^\dag\}$, we arrive at the master equation
\begin{equation}
  \ddt \rho = -\rmi [H, \rho] - \{\Gamma,\rho\} + 2\int
  \Gamma(x)\boldpsi(x)\rho\boldpsi(x)^\dag \; \rmd x,
  \label{eq:master-equation-integral}
\end{equation}
which on block form becomes
\begin{equation}
  \ddt \rho_{n} = -\rmi [H, \rho_{n}] - \{\Gamma,\rho_{n}\} + 2\int
  \Gamma(x)\boldpsi(x)\rho_{n+1}\boldpsi(x)^\dag \; \rmd x.
  \label{eq:master-equation-integral-blocks}
\end{equation}
where we have assumed an $N$-particle initial condition
$\rho(0)=\rho_N$.

We stress that the Lindblad equation (\ref{eq:master-equation-integral-blocks})
followed \emph{only} from the probability interpretation of $\|\Psi_N\|^2$ and
from the requirement that the master equation generates a Markovian,
trace-preserving and completely positive semigroup.

\section{MCTDH formulation for identical particles}
\label{sec:mctdh-for-fermions}

\subsection{The $\rho$-MCTDH manifold}
\label{sec:manifold}

We now derive the $\rho$-MCTDH approximation to the Lindblad equation
(\ref{eq:master-equation-integral}).  This method is in a sense a combination
of $\Psi$-MCTDH for identical particles
\cite{Zanghellini2003,Alon2007,Alon2008,Alon2007b}, using second quantization,
and $\rho$-MCTDH \cite{Raab1999,Raab2000}. Our method is necessarily
formulated in Fock space and describes a variable number of particles. To the
best of our knowledge, $\rho$-MCTDH has not been formulated using second
quantization in the literature. Moreover, the global variational principle
employed in $\Psi$-MCTDH (the action integral point of view) is \emph{not}
applicable for dissipative systems, since $\mathcal{D}(\rho)$ is not
self-adjoint. The second-quantization techniques used here are therefore
somewhat different from the pure-state approach. As the particular properties
of the equations of motion for the present problem are more easily exposed in
a thorough investigation, we choose to do a detailed derivation.

We repeat the Lindblad equation~(\ref{eq:master-equation-integral}) for
convenience:
\begin{align}
  \ddt \rho &= \mathcal{L}(\rho) \notag \\ 
& = -\rmi [H, \rho] + \mathcal{D}(\rho) \notag \\
& = -\rmi [H, \rho] - \{\Gamma,\rho\} + 2\int
  \Gamma(x)\boldpsi(x)\rho\boldpsi(x)^\dag \; \rmd x.
  \label{eq:master-equation-integral-rep}
\end{align}
Here, $\rho\in\TC(\mathcal{H})$ is a Fock space density operator, which (in the
exact equation) is block diagonal. $\boldpsi(x)$ destroys a particle at
the configuration-space point $x\in X$. Typically $X = \mathbb{R}^3 \times \{
\uparrow, \downarrow \}$ for a spin-$1/2$ fermion, i.e., $x=(\vec{r},m)$,
although our derivations are completely independent of $X$.

For all $t\geq 0$, $\rho(t)$ is approximated by an element in a manifold
$\mathcal{M} \subset \TC(\mathcal{H})$, with inner product inherited from
$\TC(\mathcal{M})$. (Strictly speaking, the inner product is inherited
from the Hilbert space of Hilbert-Schmidt operators, in which
$\TC(\mathcal{H})$ is dense. Not all Hilbert-Schmidt operators are
trace-class.) An approximate variational differential equation on
$\mathcal{M}$ is sought. The time-dependent variational principle
\cite{Broeckhove1988,Lubich2005,Beck2000} chooses the time derivative
$\dot{\rho} = \rmd\rho/\rmd t$ to minimize the local error in the norm induced
by $\brrakket{\cdot,\cdot}$ as follows:
\begin{equation}
  \brrakket{\delta\rho, \dot{\rho} - \mathcal{L}(\rho)} = 0, \quad \forall \delta\rho\in
  T_\rho\mathcal{M},
  \label{eq:tdvp}
\end{equation}
where $T_\rho\mathcal{M}$ is the tangent space at $\rho$, i.e., the space of
all possible time derivatives of $\rho$. Consequently, the right-hand-side
$\mathcal{L}(\rho)$ of Eq.~(\ref{eq:master-equation-integral-rep}) is
projected orthogonally onto $T_\rho\mathcal{M}$, and we have
\begin{equation*}
  \dot{\rho} = \operatornamewithlimits{argmin}_{\sigma\in
    T_\rho\mathcal{M}} \| \sigma - \mathcal{L}{\rho} \|.
 \label{eq:minimizing condition}
\end{equation*}

We choose $\mathcal{M}$ as the so-called ``type II'' density operator
manifold \cite{Raab1999,Raab2000,Meyer2009}, albeit with a slight
generalization as we consider Fock space instead of a fixed number of
particles. There is an alternate way of defining a variational
manifold, the ``type I'' manifold, but it does not reduce to the usual
MCTDH in the case of a pure state. 

The manifold $\mathcal{M}$ is defined as follows. Given a finite set $\varphi$
of $L$ single-particle functions (SPFs) $\varphi_j\in\mathcal{H}_1$, $1\leq j
\leq L$ and their corresponding creation operators $c_j^\dag$,
\begin{equation*}
  c^\dag_j = \int_X \varphi_j(x) \boldpsi(x)^\dag \; \rmd x,
\end{equation*}
we consider the subspace $\mathcal{V}_n$
of $\mathcal{H}_n$ spanned by all possible linearly
independent $n$-body
 functions built using products of $c_j^\dag$,
\begin{equation*}
  \Phi_{J[n]} = c_{j_1}^\dag c_{j_2}^\dag \cdots c_{j_n}^\dag
  \Phi_\text{vac}.
\end{equation*}
The notation $J[n]$ means an ordered tuple $(j_1,j_2,\cdots,j_n)$ of $n$
single-particle indices, i.e., $j_1\leq j_2\leq\cdots\leq j_n$. For fermions
$\Phi_{J[n]}$ is a Slater determinant and $j_1< j_2<\cdots<j_n$, while for
bosons it is a permanent. We then consider the subspace $\mathcal{V}$ of Fock
space,
\begin{equation*}
  \mathcal{V} = \bigoplus_{n=0}^N \mathcal{V}_n 
\end{equation*}
spanned by all the $\Phi_J$, where $J$ means a ordered tuple with
$n$ indeterminate. As we describe at most $N$ particles, we truncate
at $n \leq N$ in order to ensure a finite-dimensional space. (For
bosons any number of particles may occupy each $\varphi_j$, creating
an infinite dimensional space if we do not truncate the sum, even with $L$
finite.) 

Each $\rho\in\mathcal{M}$ is now defined as an arbitrary linear operator in
$\mathcal{V}$, viz, 
\begin{equation*}
  \rho = \sum_{JK} \ket{\Phi_J}B_{JK} \bra{\Phi_K}, \quad B_{JK}\in\mathbb{C}.
\end{equation*}

We see that $\rho$ is a matrix with respect to a time-dependent orthonormal
basis. We denote by $\mathbf{B}$ the matrix formed by the $B_{JK}$,
i.e., the Galerkin matrix.
To sum up, $\rho$ is parameterized in terms of an
arbitrary matrix $\mathbf{B}$ with respect to the basis generated by an arbitrary set
of $L$ SPFs $\varphi$.

The set $\varphi$ may \emph{formally} be extended to a \emph{complete} ONB
$\tilde{\varphi}$ for $\mathcal{H}_1$, a (usually infinite) set of
functions $\varphi_s$, $s>L$, such that the second-quantized
Hamiltonian is given by Eq.~(\ref{eq:ham-second-quant}), and the
CAP by Eq.~(\ref{eq:ham-with-cap}), but where the expansion
coefficients in general will depend on the particular value for
$\varphi$ and $\tilde{\varphi}$. This will be of use later on.

\subsection{Parametric redundancy and tangent space}
\label{sec:gauge}

For a given $\rho\in\mathcal{M}$, the parameters $\varphi$ and $B$ are not unique. Since $\mathcal{V}$
is determined only by the subspace spanned by $\varphi$, not the
individual $\varphi_j$, any unitary change
\begin{equation*}
  \varphi_j \longrightarrow  \sum_k \varphi_k G_{kj}, \quad G = \mathsf{U}(L)
\end{equation*}
where $\mathsf{U}(L)$ is the unitary group of $L\times L$ matrices,
yields the same space $\mathcal{V}$, and therefore the same operators
$\rho$ can be parameterized. Under the group element $G$, the basis
functions transform as
\begin{equation*}
  \Phi_{J[n]} \longrightarrow \sum_{k_1}\cdots\sum_{k_n}
  G_{k_1 j_1} \cdots G_{j_n k_n} \Phi_{K[n]} \equiv \sum_{K[N]} \!\!'\mathcal{G}_{K[n],J[n]}\Phi_{K[n]},
\end{equation*}
where the sum $\sum_{K[n]}'$ is over \emph{all} multi-indices of
length $n$, and not only ordered ones. Defining the transformation of $\mathbf{B}$ by
\begin{equation*}
  B_{JK} \longrightarrow \sum_{J'K'} \!\!' \mathcal{G}_{J'J}^* B_{J'K'} \mathcal{G}_{K'K},
\end{equation*}
we see that
\begin{equation*}
  \rho \longrightarrow \rho.
\end{equation*}
Moreover, $B_{JK}$ are all independent parameters, showing that $\mathsf{U}(L)$
is in fact the \emph{largest} group of transformations leaving $\rho$
invariant. 

The non-uniqueness of $\varphi$ and $B$ implies that given a
derivative (tangent vector) $\dot{\rho}\in T_\rho\mathcal{M}$, the
derivatives $\dot{\varphi}$ and $\dot{\mathbf{B}}$ are not unique. 
Suppose $\rho(t)\in\mathcal{M}$ is a given smooth path. There exists
$\varphi_0(t)$ and $\mathbf{B}_0(t)$ such that $\rho(t) =
\rho(\varphi_0(t),\mathbf{B}_0(t))$. By the considerations above, \emph{any} other
possible parameter path is on the form $(\varphi(t),\mathbf{B}(t)) = (\varphi_0(t)G(t),
\mathcal{G}(t)^\dag \mathbf{B}_0 \mathcal{G}(t))$, where $\varphi_0(t)G(t)$
stands for the transformation
\begin{equation*}
  \varphi_{0,j}(t) \longrightarrow \varphi_j(t) = \sum_k \varphi_{0,k}(t) G_{kj}(t).
\end{equation*}
(We consider $\varphi$ a ``matrix'' whose columns are
$\varphi_j$.) All derivatives of $\varphi_0$ are of the form
\begin{equation*}
  \dot{\varphi}_0 = \varphi_0 \eta + \chi, \quad \braket{\varphi_{0,j}|\chi_k} = 0.
\end{equation*}
The functions $\chi_j$ are all independent.
Since $GG^\dag=I_L$, we find that $\dot{G} = g G $, with $-g^\dag
= g \in \mathsf{u}(L)$, the Lie algebra of the Lie group $\mathsf{U}(L)$. The
transformed $\varphi = \varphi_0 G$ then has the derivative
\begin{equation*}
  \dot{\varphi} = \ddt{\varphi_0 G} = [\varphi_0 (\eta+g) + \chi]G,
\end{equation*}
and it is seen that if we choose $g = -\eta$, then
$\varphi$ is in fact unique, since $\dot{G} = \eta G$ uniquely
specifies $G(t)$. This is equivalent to the condition
\begin{equation*}
  \braket{\varphi_j|\dot{\varphi}_k} = 0, \quad \forall j,\;k.
\end{equation*}

In this way, there is a there is a one-to-one map between triples
$(\dot{\varphi},\dot{\mathbf{B}},g)$ and $\dot{\rho}$, with $g\in\mathsf{u}(L)$. The
element $g$ is then called a \emph{gauge choice}, and the gauge choice induces
a unique parameterization $(\varphi(t),\mathbf{B}(t))$ of $\rho(t)$. This kind
of differential geometrical structure is called a principal bundle
\cite{Lubich2008}, and is familiar in quantum field theory---but it arises in
a completely different way!

It is easily verified that
\begin{equation*}
  \ddt \Phi_J = \ddt c^\dag_{j_1}\cdots c^\dag_{j_n}\Phi_\text{vac} = D\Phi_J.
\end{equation*}
where $D$ is the operator
\begin{equation*}
  D \equiv \sum_{j=1}^L \dot{c}_j^\dag c_j
\end{equation*}
and where $\dot{c}_j^\dag$ is as the operator that creates the
single-particle function $\dot{\varphi}_j$. We observe that for
\emph{any two} single-particle functions $u$ and $v$, not necessarily
normalized, the relation
\begin{equation*}
  \{ c(u), c^\dag(v)\}_\pm \equiv c(u)c^\dag(v) \pm c^\dag(v) c(u) = \braket{u|v}
\end{equation*}
is obtained by expanding each operator in the field creation
operators. 

We are now ready to consider an arbitrary time derivative of an element
$\rho(t)\in\mathcal{M}$:
\begin{align}
  \dot{\rho} &= \sum_{J,K}  
  \ket{\dot{\Phi}_J}B_{J,K}\bra{\Phi_K} +
  \ket{\Phi_J}\dot{B}_{J,K}\bra{\Phi_K} + \ket{\Phi_J}B_{J,K}\bra{\dot{\Phi}_K}
\notag \\
&=  D\rho + \rho D^\dag + \sum_{J,K} \ket{\Phi_J}\dot{B}_{J,K}\bra{\Phi_K}. \label{eq:derivative}
\end{align}

In order to perform the projections in Eq.~(\ref{eq:tdvp}), we must
identify all linearly independent tangent vectors, i.e., all independent
admissible infinitesimal variations of $\rho$. This amounts to varying the
$B_{J,K}$ independently, and the $\varphi_j$ independently, but
according to a specific choice of gauge. For simplicity, we consider
the choice $\braket{\varphi_j|\dot{\varphi}_k}=0$, which generates the
simplest equations and are also the most common in MCTDH theory. 

From this it follows that the admissible time derivatives of
$\varphi_j$ are arbitrary functions $\vartheta = Q\vartheta$, where
\begin{equation*}
  Q \equiv I - \sum_{k} \ket{\varphi_k}\bra{\varphi_k}.
\end{equation*}
Moreover,
\begin{equation*}
  \braket{\Phi_J|\dot{\Phi}_K} = \braket{\Phi_J|D|\Phi_K} = 0, \quad \forall J, \; K.
\end{equation*}

Now, the independent variations of $\rho$ can be divided into two groups: For
each pair $J$, $K$, the matrix element $B_{J,K}$ can be changed, giving a
tangent vector
\begin{equation}
  \delta \rho = \ket{\Phi_J} \bra{\Phi_K}.
  \label{eq:delta-rho-1}
\end{equation}
An arbitrary change $\vartheta = Q\vartheta$ in $\varphi_j$ consistent with the gauge
choice gives
\begin{equation}
  \delta \rho = c^\dag(\vartheta) c_j\rho + \rho c^\dag_j c(\vartheta).
  \label{eq:delta-rho-2}
\end{equation}
Inserting these two expressions into the variational principle will yield a
complete set of differential equations for $\mathbf{B}$ in the
first case, and $\varphi$ in the latter.

\subsection{Equations of motion}
\label{sec:eom}

We use the notation $\widetilde{\sum}_j$ to indicate a sum over the
\emph{complete} set of SPFs. We let $\mathbf{H}$ be the Galerkin matrix of
$H$, i.e.,
\begin{equation*}
  H_{JK} \equiv \braket{\Phi_J|H|\Phi_K},
\end{equation*}
and analogously define $\mathbf{G}$ to be the Galerkin matrix of
$\Gamma$. We let $\mathbf{c}_j$ be the
Galerkin matrix of $c_j$, which in fact is independent of $\varphi$:
\begin{equation}
  (c_j)_{JK} = \braket{\Phi_J|c_j|\Phi_K} = \braket{\Phi'_J|c'_j|\Phi'_K}.
  \label{eq:creation-basis-invariance}
\end{equation}
The primed quantities corresponds to any other choice of SPFs.
The independence follows from the (anti-)commutator
(\ref{eq:fermion-anticomm}) which only depends on orthonormality. 

The Galerkin matrices of $H$ and $\Gamma$ can be expressed as
\begin{align}
  \mathbf{H} &= \sum_{jk} h(\varphi)_{jk} \mathbf{c}_j^\dag \mathbf{c}_k +
  \frac{1}{2}\sum_{jklm} u(\varphi)_{jklm} \mathbf{c}_j^\dag\mathbf{c}_k^\dag
  \mathbf{c}_m\mathbf{c}_l \label{eq:H-galerkin} \\
  \mathbf{G} &= \sum_{jk} \Gamma(\varphi)_{jk} \mathbf{c}_j^\dag \mathbf{c}_k \label{eq:G-galerkin}
\end{align}
The expansion coefficients are dependent on $\varphi$ at time $t$, but
the creation- and annihilation matrices are not. The Galerkin
matrices are naturally expressed using some fixed, abstract basis
due to Eq.~(\ref{eq:creation-basis-invariance}). Existing methodology
for computing matrix-vector and matrix-matrix products can be re-used
by referring to this basis.

To derive the equations of motion, we begin by inserting $\delta\rho$ from
Eq.~(\ref{eq:delta-rho-1}) into the variational principle
(\ref{eq:tdvp}). For the term $\brrakket{\delta\rho,\dot{\rho}}$ and
the term containing $[H,\rho]$, and for the sandwich term, we obtain, respectively,
\begin{widetext}
  \begin{gather}
    \brrakket{\ket{\Phi_J}\bra{\Phi_K},\sum_{J'K'} \dot{B}_{J'K'}\ket{\Phi_{J'}}\bra{\Phi_{K'}} + D\rho + \rho
      D^\dag} = \tr\left[\ket{\Phi_{K}}\bra{\Phi_{J}}\left(\sum_{J'K'}
        \ket{\Phi_{J'}}\dot{B}_{J'K'}\bra{\Phi_{K'}} + D\rho + \rho
        D^\dag \right)\right] = \dot{B}_{JK} \label{eq:rho-dot-B} \\
    \brrakket{\ket{\Phi_J}\bra{\Phi_K}, -\rmi[H,\rho]} = -\rmi \tr\left[\ket{\Phi_{K}}\bra{\Phi_{J}}\left(
        H\sum_{J'K'} \ket{\Phi_{J'}}B_{J'K'}\bra{\Phi_{K'}} - \sum_{J'K'}\ket{\Phi_{J'}}B_{J'K'}\bra{\Phi_{K'}} H
      \right) \right] = (-\rmi[\mathbf{H},\mathbf{B}])_{JK} \label{eq:commutator-B}\\
    \brrakket{\ket{\Phi_J}\bra{\Phi_K}, \widetilde{\sum_{jk}} \Gamma_{jk} c_k\rho c_j^\dag}
    = \tr\left[\ket{\Phi_{K}}\bra{\Phi_{J}}\left( 
        \widetilde{\sum_{jk}} \Gamma_{jk} c_k \sum_{J'K'} \ket{\Phi_{J'}}B_{J'K'}\bra{\Phi_{K'}}
        c_j^\dag \right) \right] = \sum_{jk} \Gamma_{jk} (\mathbf{c}_k
    \mathbf{B} \mathbf{c}_j^\dag)_{JK}. \label{eq:sandwich-B}
  \end{gather}
\end{widetext}
In Eq.~(\ref{eq:rho-dot-B}) the terms containing $D$ vanish since
$\braket{\varphi_j|\dot{\varphi}_k} = \{ c_j, \dot{c}_k^\dag \}_\pm =
0$. A similar calculation as Eq.~(\ref{eq:commutator-B}) yields
\begin{equation}
  \brrakket{\ket{\Phi_J}\bra{\Phi_K}, \{\Gamma,\rho\}} =
  (\{\mathbf{G},\mathbf{B}\})_{JK}.
  \label{eq:anticommutator-B}
\end{equation}

Assembling Eqs.~(\ref{eq:rho-dot-B})
through~(\ref{eq:anticommutator-B}), we get the equation of motion for $\mathbf{B}$:
\begin{equation*}
  \dot{\mathbf{B}} = -\rmi [\mathbf{H},\mathbf{B}] - \{
  \mathbf{G},\mathbf{B} \} + 2 \sum_{jk} \Gamma_{jk} \mathbf{c}_k
  \mathbf{B} \mathbf{c}_j^\dag.
\end{equation*}
We now make the observation that $\dot{\mathbf{B}}^\dag =
\dot{\mathbf{B}}$, showing that $\rho^\dag=\rho$ is preserved during
evolution, which we may use when we turn to the projection onto the tangent vector in
Eq.~(\ref{eq:delta-rho-2}). Let
$F$ be an arbitrary operator, and calculate
\begin{widetext}
  \begin{align*}
    \brrakket{\delta\rho, F\rho + \rho F^\dag} &= \tr\left\{[c(\vartheta)^\dag c_j \rho +
      \rho c_j^\dag c(\vartheta)](F\rho + \rho F^\dag)\right\}
    = \tr\left[c(\vartheta)^\dag c_j \rho F \rho + c(\vartheta)^\dag c_j \rho^2 F^\dag + \rho
      c_j^\dag c(\vartheta) F \rho + \rho c_j^\dag c(\vartheta) \rho F^\dag  \right] \notag \\
    &= 2 \Re \tr [c_j^\dag c(\vartheta) F \rho^2], \notag
  \end{align*}
\end{widetext}
since $c(\vartheta)\rho \equiv 0$, as $c(\vartheta)$ annihilates a function orthogonal to all the
$\varphi_j$. 

Setting $F = D = \sum_k \dot{c}_k^\dag c_k$ we obtain
\begin{align*}
    \brrakket{\delta\rho, D\rho + \rho D^\dag} &= \sum_k 2 \Re \tr [c_j^\dag c(\vartheta)
    \dot{c}_k^\dag c_k \rho^2] \notag \\ &= \sum_k 2\Re
    \braket{\vartheta|\dot{\varphi}_k}\tr(c_j^\dag c_k \rho^2),
\end{align*}
and we note in passing that $\brrakket{\delta\rho, \ket{\Phi_J}\dot{B}_{JK} \bra{\Phi_K}} =
0$, again since $c(\vartheta)\ket{\Phi_J} = 0$.

The ``sandwich'' term in the master equation also gives zero contribution, since
\begin{equation*}
  \tr\left[c(\vartheta)^\dag c_j \rho c_l \rho c_m^\dag \right] = \tr[\rho c^\dag_m c(\vartheta)^\dag
  c_j \rho] = 0,
\end{equation*}
since $c(\vartheta) c_m \ket{\Phi_J} = 0$.

In these calculations $\vartheta = Q\vartheta$ was arbitrary. Choosing $-\rmi \vartheta$ instead turns
``Re'' into ``Im'', so we may drop taking the real 
part. Assembling this, we get the equation
\begin{equation}
  \rmi \sum_k \braket{\vartheta| \dot{\varphi}_k} \tr(c^\dag_j c_k \rho^2) = \tr
  [c^\dag_j c(\vartheta) F \rho^2],
  \label{eq:spf-eq-unfinished}
\end{equation}
with $F = H - \rmi \Gamma$, and the equation must hold for all $\vartheta$ and all $j$.

We now compute the right-hand-side of
Eq.~(\ref{eq:spf-eq-unfinished}) for arbitrary single-particle and
two-particle operators $F$. Suppose at first
\begin{equation*}
  F = \widetilde{\sum_{jk}} f_{jk} c^\dag_j c_k, \quad f_{jk} =
  \braket{\varphi_j|f|\varphi_k}.
\end{equation*}
Upon insertion in Eq.~(\ref{eq:spf-eq-unfinished}), we find
\begin{align*}
  \tr [c^\dag_j c(\vartheta) F \rho^2] &= \widetilde{\sum_{kl}} f_{kl} \tr[c^\dag_j c(\vartheta)
   c^\dag_k c_l \rho^2] \\
  &= \widetilde{\sum_k}\sum_l
  \braket{\vartheta|\varphi_k}\braket{\varphi_k|f|\varphi_l} \tr(c^\dag_j c_l \rho^2)
  \\
  &= \sum_l
  \braket{\vartheta|f|\varphi_l} \tr(c^\dag_j c_l \rho^2)
\end{align*}
In the last calculation, we used
$\widetilde{\sum}_j\ket{\varphi_j}\bra{\varphi_j} = I$. 

Second, suppose $F$ is a two-particle operator, viz,
\begin{equation*}
  F = \frac{1}{2}\widetilde{\sum_{jklm}} f_{jklm} c^\dag_j c^\dag_k c_m c_l, \quad f_{jklm} =
  \braket{\varphi_j\varphi_k|f|\varphi_l\varphi_m},
\end{equation*}
where we assume $f_{jklm} = f_{kjml}$. The inner product is on
$\mathcal{H}_1\otimes \mathcal{H}_1$, i.e., the brackets are not
antisymmetrized. Now the right-hand-side of
Eq.~(\ref{eq:spf-eq-unfinished}) becomes
\begin{widetext}
\begin{align*}
  \tr [c^\dag_j c(\vartheta) F \rho^2] &= \frac{1}{2}\widetilde{\sum_{klmn}} f_{klmn} \tr[c^\dag_j c(\vartheta)
   c^\dag_kc^\dag_l c_n c_m \rho^2] 
  = 
  \frac{2}{2}\widetilde{\sum_k}\sum_{lmn}
  \braket{\vartheta|\varphi_k}\braket{\varphi_k\varphi_l|f|\varphi_m\varphi_n}
  \tr(c^\dag_jc^\dag_l c_n c_m \rho^2)
  \\
  &= \sum_{lmn}
  \braket{\vartheta\varphi_l|f|\varphi_m\varphi_n} \tr(c^\dag_jc^\dag_l c_n c_m \rho^2)
\end{align*}
\end{widetext}
In the first step, we used the symmetry of $f_{klmn}$ and the (anti-)commutator
relation for the creation operators operators, the result being the
same regardless of particle statistics.

Three-particle operators, or even higher, are computed in similar fashion. For
a three-body operator
\begin{equation*}
  F=\frac{1}{3!}\widetilde{\sum_{jkl}}\widetilde{\sum_{pqr}} f_{jklpqr} c^\dag_j
  c^\dag_k c^\dag_l c_r c_q c_p
\end{equation*}
we obtain the right-hand side
\begin{equation*}
  \tr [c^\dag_j c(u) F \rho^2] 
  = \sum_{klpqr} \braket{\vartheta\varphi_k\varphi_l|f|\varphi_p\varphi_q\varphi_r}
  \tr(c^\dag_jc^\dag_k c^\dag_l c_r c_q c_p \rho^2)
  \label{eq:three-body-2}
\end{equation*}
In all cases, the combinatorial factor cancels due to symmetry properties.

The matrix 
\begin{equation}
  S_{jk} \equiv \tr(c^\dag_j c_k \rho^2) = \tr(\mathbf{c}^\dag_j
  \mathbf{c}_k \mathbf{B}^2)
  \label{eq:one-body-reduced}
\end{equation}
defines the $\rho$-MCTDH analogue of the \emph{reduced
  one-body density matrix} entering at the same location in standard $\Psi$-MCTDH theory. Similarly,
the analogue of the reduced
two-body density matrix is defined by
\begin{equation}
  S^{(2)}_{jklm} \equiv \tr( c^\dag_j c^\dag_k c_m c_l \rho^2) = \tr(\mathbf{c}^\dag_j\mathbf{c}^\dag_k
  \mathbf{c}_m\mathbf{c}_l \mathbf{B}^2),
  \label{eq:two-body-reduced}
\end{equation}
and so on.

We may now assemble the various one- and two-body contributions to the
SPF equation of motion:
\begin{equation*}
  \rmi \sum_{k} \braket{\vartheta|\dot{\varphi}_k} S_{jk} =  \sum_k \braket{\vartheta|(h - \rmi \Gamma)|\varphi_k} S_{jk}
  + \sum_{klm} \braket{\vartheta\varphi_k|u|\varphi_l\varphi_m} \varphi_l S^{(2)}_{jklm},
\end{equation*}
which holds for \emph{all} $\vartheta=Q\vartheta$. Since
$\dot{\varphi}_j=Q\dot{\varphi}_j$, we arrive at the final
single-particle equations of motion:
\begin{align*}
  \rmi \sum_{k} \dot{\varphi}_k S_{jk} &= Q \sum_k (h - \rmi \Gamma)\varphi_k S_{jk}
  + \sum_{klm} Q \braket{\;\cdot\; \varphi_k|u|\varphi_l \varphi_m} \varphi_l S^{(2)}_{jklm} \notag \\
  &= Q \sum_k (h - \rmi \Gamma)\varphi_k S_{jk}
  + \sum_{klm} Q U_{km} \varphi_l S^{(2)}_{jklm},
\end{align*}
where
\begin{equation*}
\braket{\;\cdot\; \varphi_k|u|\varphi_l \varphi_m} \equiv \int
\overline{\varphi_k(y)} u(x,y) \varphi_l(x)\varphi(y) \; \rmd y,
\end{equation*}
and where the mean-field potentials $U_{km}$ are defined by
\begin{equation}
  U_{km}(x) \equiv \int \overline{\varphi_k(y)}u(x,y)\varphi(y) \; \rmd y.
  \label{eq:mean-field}
\end{equation}
Assuming $u(x,y)$ to be a local potential, $U_{km}(x)$ is also a local
one-body function.

\subsection{Discussion}
\label{sec:eom-discussion}

Let us sum up the equations of motion for the density operator $\rho$. The
Galerkin matrix elements $B_{J,K}$ evolves according to
\begin{equation}
  \dot{\mathbf{B}} = -\rmi [\mathbf{H},\mathbf{B}] - \{
  \mathbf{G},\mathbf{B} \} + 2 \sum_{jk} \Gamma_{jk} \mathbf{c}_k \mathbf{B}
  \mathbf{c}_j^\dag, \label{eq:B-eom}
\end{equation}
while the SPFs evolve according to
\begin{equation}
\rmi \sum_{k} \dot{\varphi}_k S_{jk} = Q \sum_k (h - \rmi \Gamma)\varphi_k S_{jk}
  + \sum_{klm} Q U_{km} \varphi_l S^{(2)}_{jklm}, \label{eq:Phi-eom}
\end{equation}
where $S_{jk}$, $S^{(2)}_{jklm}$ and $U_{km}(x)$ were defined in
Eqs.~(\ref{eq:one-body-reduced}), (\ref{eq:two-body-reduced}) and
(\ref{eq:mean-field}), respectively.

Eq.~(\ref{eq:Phi-eom}) is virtually identical to the standard $\Psi$-MCTDH
equation of motion for the SPFs. The only difference lies in the definitions
of $S$ and $S^{(2)}$. As for Eq.~(\ref{eq:B-eom}), we see that the main
difference lies in the evolution of a \emph{matrix} $\mathbf{B}$ instead of a
coefficient vector.

Eq.~(\ref{eq:Phi-eom}) is typically discretized using discrete-variable
representation (DVR) techniques, FFT methods, finite differences or similar
\cite{Meyer2009}, and the single-particle operator $h-\rmi\Gamma$ is then
represented correspondingly. Equivalently, the single-particle space
$\mathcal{H}_1$ is approximated by the finite-dimensional space dictated by
the discretization, inducing a finite-dimensional Fock space to begin
with. Note that even though $h-\rmi\Gamma$ is non-Hermitian, orthonormality of
$\varphi$ is conserved during evolution.

Like standard $\Psi$-MCTDH, the matrix $S$ needs to be inverted to
evaluate the SPF differential equation; a well-known issue with MCTDH-type
methods. It may happen that $S$ becomes singular for some reason, in which
case a regularization approach is needed \cite{Meyer2009}. In most
applications, this happens very rarely; typically at $t=0$ due to the choice
of initial conditions, but experience suggests it does not affect the final
results. This is, however, not trivial from a mathematical point of view, and
for the sake of definiteness in the present work, we shall check that $S$ is
non-singular for our numerical experiment in
Sec.~\ref{sec:experiments}.

Eq.~(\ref{eq:B-eom}) should be compared with the original Lindblad
equation~(\ref{eq:master-eq-second-quant}). Also, if $\dot{\varphi}\equiv 0$, we
obtain the variational equation of motion in a fixed linear basis, i.e., what
is obtained using a full configuration-interaction type approach. However, the
Galerkin matrices defined in Eqs.~(\ref{eq:H-galerkin}) and
(\ref{eq:G-galerkin}) have time-dependent coefficients $h_{jk}$, $u_{jklm}$,
and $\Gamma_{jk}$ which must be computed along the flow. This is a non-trivial
task in general, and techniques common for $\Psi$-MCTDH can be employed to
deal with this in approximate ways \cite{Caillat2005,Meyer2009}.

Note that $\mathbf{B}$ retains the natural block structure with respect to the
number of particles,
cf.~Eq.~(\ref{eq:master-eq-second-quant-blocks}). Eq.~(\ref{eq:B-eom})
can be written as
\begin{equation*}
  \dot{\mathbf{B}}_n = -\rmi [\mathbf{H},\mathbf{B}_n] - \{
  \mathbf{G},\mathbf{B}_n \} + 2 \sum_{jk} \Gamma_{jk} \mathbf{c}_k \mathbf{B}_{n+1}
  \mathbf{c}_j^\dag 
\end{equation*}
and this is the most memory-economical representation, since the off-diagonal
blocks vanish if $\rho(0)$ is a pure state with $N$ 
particles. In that case, $\mathbf{B}_N$ can furthermore be represented by a vector
$\Psi_N\in\mathcal{V}_N$ at all times, and there is no need to propagate the full
block. Due to the presence of \emph{all} the blocks in the definition of $S$
and $S^{(2)}$, however, this pure state cannot be evolved with $\Psi$-MCTDH independently of the other blocks.

If the dissipative terms vanish (i.e., if
$\Gamma=0$), and if $\rho(0)$ is a pure state, the evolution is easily seen to
be equivalent to a $\Psi$-MCTDH calculation.  In $\rho$-MCTDH, $\tr(\rho)$ is
not in general conserved \cite{Meyer2009}, but it is so for closed systems, i.e.,
when $\mathcal{D}(\rho)\equiv 0$. However, it is easily checked in the present
case that, indeed,
\begin{equation*}
  \ddt \tr(\rho) = \ddt \tr(\mathbf{B}) = 0.
\end{equation*}
Also, energy $\tr(H\rho)$ is exactly conserved if $\tr(\Gamma\rho)=0$, that is
to say, whenever the system does not touch the CAP.

For the actual implementation of the evolution equations, it is useful to
employ a generic enumeration scheme for the many-body basis states. In
many-body codes, the Galerkin matrices (other than $\mathbf{B}$) are rarely
constructed in memory; instead the single- and double-particle integrals are kept in memory and the
explicit action of ${H}$ is computed using Eq.~(\ref{eq:H-galerkin}), for
which the action of $\mathbf{c}_j$ and $\mathbf{c}_j^\dag$ are implemented for
example via mapping techniques as suggested in Ref.~\cite{Streltsov2010}; or
simply using binary integers to represent a fermion state, and bitwise
manipulations to define the action of $\mathbf{c}_j$ etc.; a common technique
in many-body nuclear physics calculations \cite{Whitehead1977}.

As for choosing intitial conditions, we observe that as a
generalization of $\Psi$-MCTDH capable of treating particle loss, a
pure state $\rho(0)=\ket{\Psi_N}\bra{\Psi_N}$ will be the usual choice. In
that case, experience from $\Psi$-MCTDH can be applied
\cite{Meyer2009}. Typical choices are single determinants/permanents or
stationary states computed by imaginary time propagation, or combinations
thereof as in the numerical experiment below.

\section{Numerical experiment}
\label{sec:experiments}

We present a numerical experiment for a model problem consisting of
spin-polarized fermions in one spatial dimension. We will study a situation
where the initial state is a pure state with $N=3$ particles whose norm gradually
decreases due to a CAP. The situation is similar to the study in
Ref.~\cite{Nest2005}. 

\begin{figure}
  \includegraphics{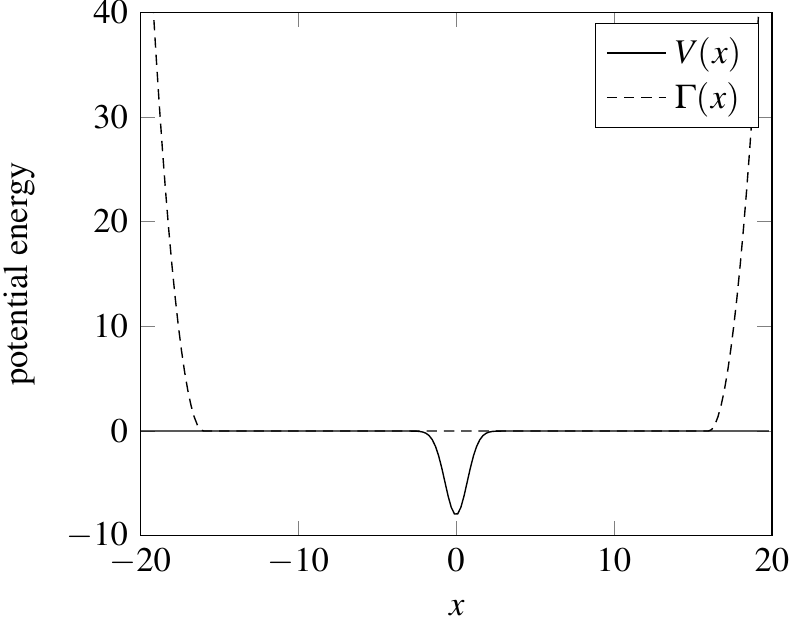}
  \caption{Single-particle trap potential $V(x)$ and complex absorbing
    potential $\Gamma(x)$ used in the numerical experiment. The trap potential is
    Gaussian, see Eq.~(\ref{eq:trap}), and the absorber is
    quadratic outside $[-16,16]$, see Eq.~(\ref{eq:absorber}).\label{fig:model}}
\end{figure}

We truncate the domain $\mathbb{R}$ to $[-R,+R]$, where $R=20$.  The
single-particle Hamiltonian of our model is
\begin{equation*}
  h = T + V(x) = -\frac{1}{2}\frac{\partial^2}{\partial x^2} + V(x)
\end{equation*}
where the one-body potential is of Gaussian shape
\begin{equation}
  V(x) = -8 \exp[-1.25 x^2].
  \label{eq:trap}
\end{equation}
Numerically we find that $V(x)$ supports $4$ bound one-body
states. We choose a very simple CAP of standard power-form:
\begin{equation}
  \Gamma(x) = \theta(|x|-R')(|x|-R')^2 ,
  \label{eq:absorber}
\end{equation}
where $\theta(x)$ is the Heaviside function. The particles are unaffected by
the CAP in the region $[-R',R']$, where we we set $R'=16$. We have verified
that in the energy ranges of the calculations, very little reflection or
transmission is generated by $\Gamma$. Fig.~\ref{fig:model} shows the Gaussian
well and the absorber.

The particles interact via a smoothed Coulomb potential,
\begin{equation*}
  u(x_1,x_2) = 2 [(x_1-x_2)^2 + 0.1^2]^{-1/2},
\end{equation*}
which is long-ranged.

For discretizing the one-body space, we choose the standard fast Fourier
transform based method with $N_\text{grid}=128$
equidistant points with spacing $\Delta x = 2R/{N_\text{grid}}$
\cite{Tal-Ezer1984}. 

For propagating the master equation, we choose a variational splitting scheme
\cite{Lubich2004,Lubich2008}, propagating the equations of motion with $H'=T$, i.e.,
kinetic energy only for a time step $\tau/2$, and then $H' = H-T$ for a time
step $\tau$, and finally $H'=T$ for a time step $\tau/2$ again. This
constitutes the propagation of a complete time step $\tau$. While being
simple, the scheme has the advantage of having local error
$\mathcal{O}(\tau^3)$, that the $T$-propagation is numerically exact, and that
the time step is not restricted to be $\tau = \mathcal{O}(\Delta x^2)$. The
potential step is integrated using a standard explicit fourth-order
Runge-Kutta method for simplicity, which is sufficient for our purposes.

The initial condition is chosen as follows. Let $\Psi_2(x_1,x_2)$ be the
two-body ground state of the CAP-less Hamiltonian. This is computed
numerically by propagating the \emph{standard} $\Psi$-MCTDH equations in
imaginary time $t = -\rmi s$ using $L=4$ single-particle states. It follows,
that this state is also a stationary for the present $\rho$-MCTDH method with
a CAP as long as the overlap with the CAP is negligible. We have checked that
this is indeed the case: Propagating the master equation with the two-body
state as initial condition leads to an absorption probability of $2.7\times
10^{-10}$ at $t=t_\text{final} = 30$ which can safely be ignored.

We act upon $\Psi_2$ with a creation operator $c(g)^\dag$, where $g(x)$ is a
Gaussian of the form
\begin{equation*}
  g(x) = Q C \exp[-(x+2)^2/0.75 + \rmi 3 x],
\end{equation*}
where $Q$ projects away the $4$ SPFs in the initial condition. $g(x)$
describes an incoming particle of momentum $k=3$ starting out at $x_0=-2$. The
final three-body initial state is then
\begin{equation*}
  \rho(0)= c^\dag(g) \ket{\Psi_2}\bra{\Psi_2}c(g).
\end{equation*}
The initial $\varphi$ then consists of the $L=5$
functions consisting of the $4$ SPFs from the ground-state
computation, and the single state $g(x)$.

\begin{figure}
  \includegraphics{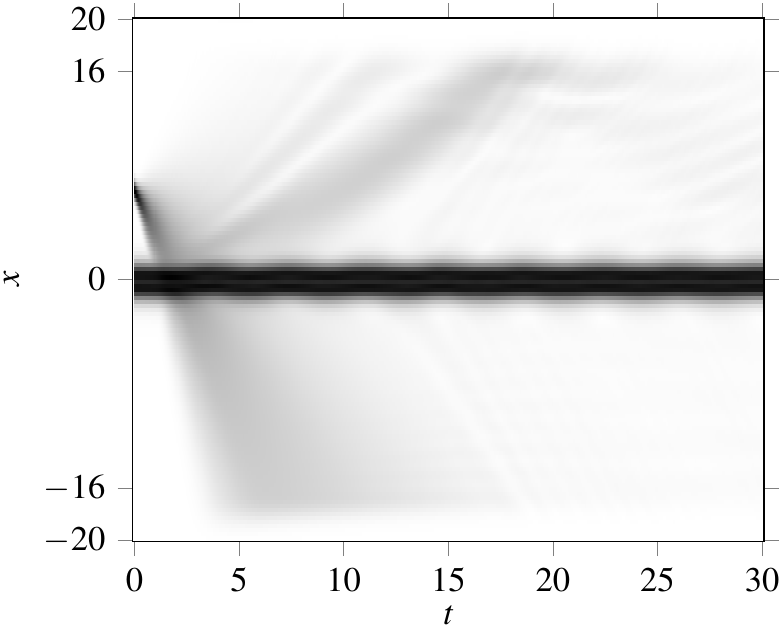}
  \caption{Space-time graph of square root $\sqrt{n(x,t)}$ of particle
    density in numerical experiment. Darker areas have higher
    density. The square root enhances contrast, but exaggerates low
    densities. From the plot, we can see that a single-particle
    function of Gaussian form is scattered off a bound two-particle
    state in a Gaussian well. The reflected and transmitted parts are
    absorbed by the CAP, revealing an oscillating trapped function of
    fewer particles.\label{fig:density-plot}}
\end{figure}

Using this initial condition, we propagate $\rho(t)$ for $t \leq
t_\text{final}=30$. Fig.~\ref{fig:density-plot} shows a space-time
graph of the particle density $n(x,t)$ given by
\begin{equation*}
  n(x,t) \equiv \tr [\boldpsi^\dag(x)\boldpsi(x)\rho(t)].
\end{equation*}
As expected, the plot shows the initial advance of the Gaussian
wavepacket and its scattering off the well and the two-particle ground
state. It is seen that scattering occurs both in the forward and backward
direction. The scattered probability is absorbed upon entering
the region $|x| \leq R'$, and a system composed of
less than three particles is seen to remain. Superficially, it is an
oscillating two-particle system. The system's energy is $E \approx -7.355$.

\begin{figure}
  \includegraphics{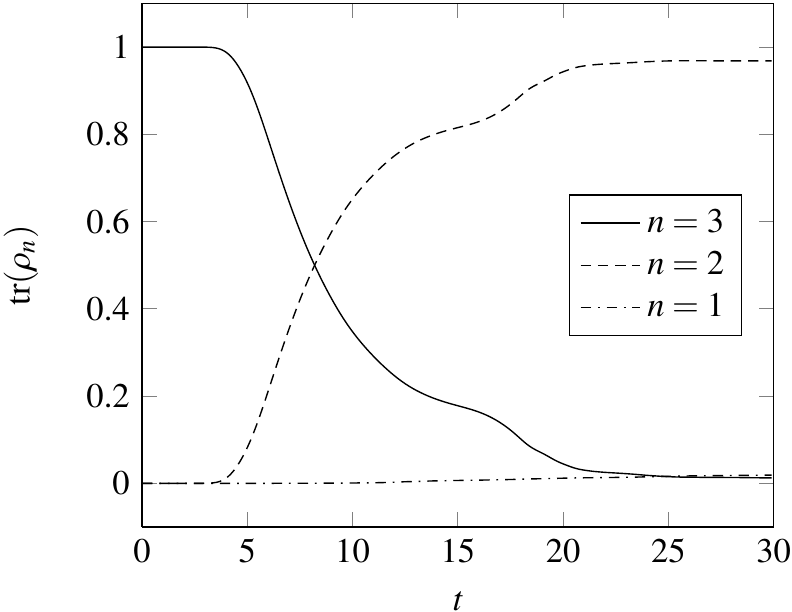}
  \caption{Plot of the probabilities $p_n=\tr(\rho_n)$ of having $n$
    particles in the system as function of $t$. The absorption events
    seen in Fig.~\ref{fig:density-plot} is seen to correspond to
    marked changes in the probabilities. $p_0$ is too small to merit
    an interesting plot.\label{fig:trace-plot}}
\end{figure}

However, the process is more complex, and by computing the probabilities
$p_n(t) = \tr[\rho_n(t)]$ of having $n$ particles in the system we may see
what happens in more detail.  In Fig.~\ref{fig:trace-plot} $p_n$ is plotted
for each $1\leq n \leq 3$. As the scattered probability density is absorbed,
the probability of having $n=3$ particles decreases and the probability of
$n=2$ increases correspondingly. However, especially the absorption of the
back-scattered wave reveals something interesting: the probability of having
$n=1$ particle in the system clearly becomes significant in this process: the
bound two-particle system has a significant probability of being ionized by
the collision, leaving a single particle. By inspecting the probability
density $n_1(x,t) = \tr[\boldpsi^\dag(x)\boldpsi(x)\rho_1(t)]$ we verify that
it corresponds to a bound one-body state superimposed on the two-body
state. The probability $p_0 \approx 3.43 \times 10^{-4}$ at
$t=t_\text{final}$, showing a very small probability of all particles
vanishing. It is therefore not plotted.

Although the initial bound two-particle state had negligible overlap with the
CAP, there may still be errors introduced by a placing the CAP too close to the
interacting system. For example, if a particle is absorbed prematurely, the
remaining system may miss some correlations. Moreover, there seems to be a
finite remaining probability of having three particles in the system. This is
most likely due to reflections off or transmissions through the non-ideal
absorber $\Gamma$ (which only was chosen for illustrative purposes), and not a
bound three-body state. We have not investigated this in detail for the
present experiment.

\begin{figure}
  \includegraphics{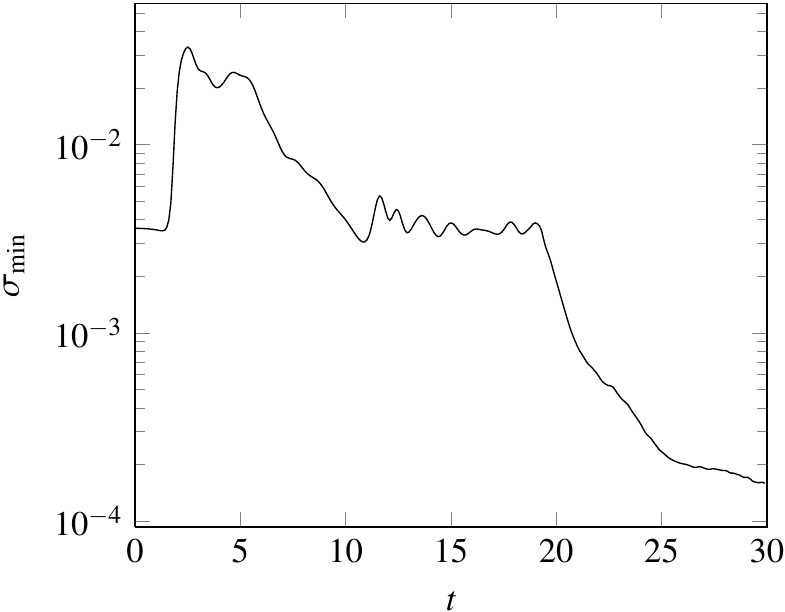}
  \caption{Smallest eigenvalue of the matrix $S$ which needs to be
    inverted at each $t$. As one particle is almost entirely absorbed,
    the smallest eigenvalue falls off rapidly. It peters out at
    $\approx 1.57\times 10^{-4}$, safely away from zero.\label{fig:svd-plot}}
\end{figure}

Finally, we address the non-singularity of the matrix $S$ in
Eq.~\eqref{eq:Phi-eom}. In Fig.~\ref{fig:svd-plot} the smallest
eigenvalue of $S$ is plotted as function of $t$. At the very last leg of the
evolution, this eigenvalue $\sigma_\text{min}$ drops off quickly.

A sharp fall-off of $\sigma_\text{min}$ is to be expected when a
particle is almost entirely absorbed. This may be understood in terms
of a non-interacting system. If interactions are not present, the
whole system may be described by the SPFs alone, each evolving
according to the non-Hermitian Schrödinger equation under the
single-particle Hamiltonian $h - \rmi \Gamma$. Thus, the eigenvalues
$\sigma_j$ only change because of the CAP. As a particle is
absorbed, one eigenvalue goes to zero.

Usually, small eigenvalues $\sigma_j$ give very rapidly changing
SPFs. However, we have observed that the natural orbital corresponding to
$\sigma_\text{min}$, which we observe resides in the CAP region and thus
represents the absorbed particle, does not change significantly in the last
leg of the evolution. This indicates that the SPF no longer becomes relevant
for the description in in the sense that the right-hand side of
(\ref{eq:Phi-eom}) de-couples from this SPF.  In this way, the
near-singularity of $S$ stemming from particle absorption may in fact have no
impact at all on the evolution. This may be related to the simple fact that
for a pure two-fermion system, the single-particle reduced density matrix is
singular whenever $L$ is an odd number. (It is easy to show that the
eigenvalues are zero or of multiplicity 2 in the two-fermion case.) We cannot
draw any firm conclusions concerning this from our simple experiment, except
for pointing out that $S$ becoming singular may have different causes and
consequences compared to pure state-MCTDH.

\section{Mixtures of species}
\label{sec:bosons-and-mixtures}

The density operator MCTDH formulation for identical fermions is readily
generalized to mixtures of arbitrary number of species of particles, such as
mixtures of ${}^3$He and ${}^4$He (fermions and bosons). Using
second quantization, the derivation becomes analogous to the treatment in
Sec.~\ref{sec:mctdh-for-fermions} and Ref.~\cite{Alon2007b}, so we
only state the main results here.

As Ref.~\cite{Alon2007b} we consider two different species of particles $A$
and $B$ for simplicity, as the generalization to $K$ species follows
immediately. Each species have a Fock space $\mathcal{H}^{(i)}$, $i =
{A,B}$. The total Hilbert space is the product space
\begin{equation*}
  \mathcal{H} = \mathcal{H}^{(A)} \otimes \mathcal{H}^{(B)}.
\end{equation*}
Each species is assigned a set $\varphi^{(i)}$ of single-particle
states $\varphi^{(A)}(x)$ and $\varphi^{(B)}(y)$, but they have no \emph{a
  priori} connection, as the single-particle spaces may be very
different. Consequently, the operators $a_j^{(\dag)}$ (for species $A$) and
$b_k^{(\dag)}$ (for species $B$) all commute since the species are
distinguishable from each other. As previously, the creation
operators are used to construct finite-dimensional Fock spaces
$\mathcal{V}^{(i)}$ with determinant or permanent basis functions
$\Phi^{(i)}_J$. For the product space, the basis functions are 
\begin{equation*}
  \Phi^{(A)}_{J[n]}\otimes\Phi^{(B)}_{K[m]} = a^\dag_{j_1}\cdots a^\dag_{j_n}
  b^\dag_{k_1}\cdots b^\dag_{k_m} \Phi_\text{vac}.
\end{equation*}
Note that as the particles are distinguishable, we speak of
$(n,m)$-particle states. Fock space is divided into subspaces with $n$
particles of species $A$, and $m$ particles of species $B$. The
density operator $\rho$ will then be block diagonal with respect to
the particle numbers:
\begin{equation*}
  \rho = \sum_{n=0}^{N_A} \sum_{m=0}^{N_B} \rho_{n,m}
\end{equation*}
where $N_i$ are the maximum number of particles in the system,
determined by the initial condition.

Each species has its internal Hamiltonian, but for the equations not
to separate into the previously studied case, we need an
interaction. A generic two-body inter-species interaction may be
written as
\begin{equation*}
  W = \sum_{i=1}^{n}\sum_{j=1}^{m} w(x_i, y_j) = \sum_{j,l=1}^{L^{(A)}} \sum_{k,m=1}^{L^{(B)}} w_{jklm}
  a^\dag_j a_l b^\dag_k b_m 
\end{equation*}
in first and second quantization form, respectively. Here,
\begin{equation*}
  w_{jklm} = \braket{\varphi^{(A)}_j \varphi^{(B)}_k | w(x,y) |
    \varphi^{(A)}_l \varphi^{(B)}_m}.
\end{equation*}
The usual factor $1/2$ is not present, since the particles are not identical.

Each species also has its own absorber $\Gamma^{(i)}$, which need not
have any \emph{a priori} relation.

Working through the equations of motion, noting that each species'
SPFs are independent from each other, we obtain the following equation
for the Galerkin matrix blocks $\mathbf{B}_{n,m}$:
\begin{align*}
  \dot{\mathbf{B}}_{n,m} &= -\rmi[\mathbf{H},\mathbf{B}_{n,m}] - \{\mathbf{G}^{(A)}
  + \mathbf{G}^{(B)}, \mathbf{B}_{n,m} \} \\
  &+ 2\sum_{jk} \Gamma^{(A)}_{jk} \mathbf{a}_k \mathbf{B}_{n+1,m}
  \mathbf{a}_j^\dag + 2\sum_{jk} \Gamma^{(B)}_{jk} \mathbf{b}_k \mathbf{B}_{n,m+1}
  \mathbf{b}_j^\dag
\end{align*}
with $\mathbf{H} = \mathbf{H}^{(A)} +  \mathbf{H}^{(B)} +  \mathbf{W}$
and an otherwise obvious notation.

We obtain an SPF equation of motion for each species. They will contain
species-specific analogues $S^{(i)}$ of $S$, and of the mean fields  $U^{(i)}$
of $U$, and also \emph{inter species} analogues of the reduced 
two-body density matrix elements $S^{(AB,2)}$ and mean fields due to $W$,
exactly as in the $\Psi$-MCTDH \cite{Alon2007b}. These are
defined by
\begin{align*}
  S^{(AB,2)}_{jklm} &\equiv S^{(BA,2)}_{kjml} \equiv \tr(\rho^2 a^\dag_j a_l b^\dag_k b_m), \\
  W^{(A)}_{km} &\equiv \int \overline{\varphi^{(B)}(y)} w(x,y) \varphi^{(B)}(y) \;
  \rmd y, \intertext{and}
  W^{(B)}_{km} &\equiv \int \overline{\varphi^{(A)}(x)} w(x,y) \varphi^{(A)}(x) \;
  \rmd x,
\end{align*}
respectively. 
We get the SPF equations of motion
\begin{widetext}
\begin{align*}
\rmi \sum_{j} \dot{\varphi}_j^{(A)} S^{(A)}_{kj} &= Q^{(A)} \Big[\sum_j
(h^{(A)} - \rmi \Gamma^{(A)})\varphi_j^{(A)} S_{kj}^{(A)}  
  +  \sum_{klm}  U^{(A)}_{km} \varphi_l^{(A)}
  S^{(A,2)}_{jklm} 
  + \sum_{klm} W^{(A)}_{km} \varphi^{(A)}_l S^{(AB,2)}_{jklm}
  \Big] 
\\
\rmi \sum_{j} \dot{\varphi}_j^{(B)} S^{(B)}_{kj} &= Q^{(B)} \Big[\sum_j
(h^{(B)} - \rmi \Gamma^{(B)})\varphi_j^{(B)} S_{kj}^{(B)}  
  +  \sum_{klm}  U^{(B)}_{km} \varphi_l^{(B)}
  S^{(B,2)}_{jklm} 
  + \sum_{klm} W^{(B)}_{km} \varphi^{(B)}_l S^{(AB,2)}_{lmjk} \Big] .
\end{align*}
\end{widetext}
These equations have an obvious symmetry with respect to particle
species interchange. These equations are identical to those given in
Ref.~\cite{Alon2007b}, except for the reduced density matrices being
defined in terms of a density operator and not a pure state.

The generalization to $K$ species is straightforward, and we refrain from
going into further detail. Note however, that one interesting special case is
obtained when the number of species equals the number of initial particles, so
that the initial state is a density operator in the space
\begin{equation*}
  \mathcal{H} = \mathcal{H}^{(1)}_1\otimes \mathcal{H}^{(2)}_1 \otimes
  \cdots \otimes \mathcal{H}^{(K)}_1.
\end{equation*}
In this case it seen that, as all particles in the system are in fact
distinguishable, we have obtained the usual MCTDH method for density
operators, but with a CAP. Of course, the same can be said of pure-state MCTDH
for mixtures of particles -- as we approach $K=N$ species, where $N$ is the
number of particles, we are back at plain MCTDH for distinguishable particles,
and the circle is closed: MCTDH for identical particles can be viewed as MCTDH
with (anti-)symmetry constraints on the coefficients, and plain MCTDH can be
viewed as $N$-species MCTDH for mixtures.

\section{Conclusion}
\label{sec:conclusion}

A system of $N$ particles described by a Hamiltonian with a complex
absorbing potential evolves irreversibly in time. In order to describe
the remaining particles as some are lost to the absorber, a master
equation on Lindblad form in Fock space is needed, as first
demonstrated in Ref.~\cite{Selsto2010}. This equation was discussed at
length, and a multiconfigurational time-dependent Hartree method was
presented that is a strict generalization of standard pure-state MCTDH
evolution for identical particles or mixtures. A numerical experiment
on a simple system of $N=3$ spin-polarized fermions was reported.

\acknowledgments 

The author wishes to thank Dr.~Sølve Selstø for fruitful discussions.
This work is supported by the DFG priority programme
SPP-1324. Further financial support of the CMA, University of Oslo,
is gratefully acknowledged.


\end{document}